\documentclass[fleqn,twoside]{article}
\usepackage[headings]{espcrc2}

\readRCS
$Id: espcrc2.tex,v 1.2 2004/02/24 11:22:11 spepping Exp $
\usepackage{graphicx}
    \usepackage{amssymb}
     \usepackage{pifont}
\usepackage[figuresright]{rotating}
 \usepackage{bm}

\newcommand{\AmS}{{\protect\the\textfont2
  A\kern-.1667em\lower.5ex\hbox{M}\kern-.125emS}}
\newcommand{\Ds}{\displaystyle}
\usepackage{color}

   \newcommand{\RedTn}[1]{\textcolor{red}{#1}}

\title{{\hfill RUB-TPII-09/09 }\\ [1cm]
Two-loop contribution to the pion transition form factor
vs. experimental data}

\author{S.~V.~Mikhailov\address[BLTP,JINR]{Bogoliubov Laboratory of Theoretical
Physics, JINR, 141980 Dubna, Russia\\
        }%
        \thanks{Talk presented by the first author at the 3rd Joint International
        Hadron Structure '09 Conference, Tatransk\'a \v{S}trba, Slovakia,
        August 30th - September 3rd, 2009.},
        N.~G.~Stefanis \addressmark[BLTP,JINR]{}%
\thanks{Institut f\"{u}r Theoretische Physik II,
        Ruhr-Universit\"{a}t Bochum, D-44780 Bochum, Germany}
}

\runtitle{Two-loop results for pion-photon transition vs. data}
\runauthor{S.~V.~Mikhailov and N.~G.~Stefanis}

\begin{document}

\begin{abstract}
We present predictions for the pion-photon transition form factor,
derived with the help of light-cone sum rules and including
the main part of the NNLO radiative corrections.
We show that, when the Bakulev-Mikhailov-Stefanis (BMS) pion
distribution amplitude is used, the obtained predictions agree well
with the CELLO and the CLEO data.
We found that no model distribution amplitude can reproduce the
observed $Q^2$ growth of the new BaBar data, though the BMS model
complies with several BaBar data points.
\vspace{1pc}
\end{abstract}

\maketitle

\section{Form factor $\mathbf{F^{\gamma^{*}\gamma^{*}\pi}}$ in QCD,
         collinear factorization}
\label{sec:col-fac}

This transition form factor describes the process
$\gamma^*(q_1)\gamma^*(q_2)\to \pi^0(p)$
and is given by the following matrix element
($-q_{1}^2\equiv Q^2>0, -q_2^2\equiv q^2\geq 0$)
\begin{eqnarray}
  \int d^{4}x e^{-iq_{1}\cdot z}
  \langle
         \pi^0 (p)\mid T\{j_\mu(z) j_\nu(0)\}\mid 0
  \rangle
=&& \nonumber \\
  i\epsilon_{\mu\nu\alpha\beta}
  q_{1}^{\alpha} q_{2}^{\beta}
  \cdot F^{\gamma^{*}\gamma^{*}\pi}(Q^2,q^2)\, .
&& \label{eq:matrix-element}
\end{eqnarray}
For sufficiently large photon momenta
$Q^2, q^2 \gg m_\rho^2$ (where the  hadron scale is set by the
$\rho$-meson mass $m_\rho$), all binding effects can be accumulated
into a universal (twist-two) pion distribution amplitude
(DA), so that, on account of collinear factorization, one obtains
the form factor as a convolution:
\begin{eqnarray}
  F^{\gamma^{*}\gamma^{*}\pi}(Q^2,q^2)
\! &\!\!\!\!\!\!\!\!\!\!=\!\!\!\!\!\!\!\!\!\!& \!
  T(Q^2,q^2,\mu^2_{\rm F};x)
\otimes
  \varphi^{(2)}_{\pi}(x;\mu^2_{\rm F})
\nonumber \\
&& \!\!\! + \ O\left( Q^{-4} \right) \ .
\label{eq:convolution}
\end{eqnarray}
While the pion DA $\varphi^{(2)}_{\pi}$, which represents a
parameterization of the pion matrix element at the (low) factorization
scale $\mu^2_{\rm F}$, cannot be calculated from first principles and
has to be modeled within some nonperturbative approach, the amplitude
$T$, describing the hard parton subprocesses, can be calculated in QCD
perturbation theory.
On the other hand, also the evolution of the pion DA with
$\mu^2_{\rm F}$ is controlled by perturbatively calculable evolution
kernels and associated anomalous dimensions.
Hence, in leading order (LO) of the strong coupling and at leading
twist two one has
 \begin{eqnarray}
   F^{\gamma^{*}\gamma^{*}\pi}
\! = \!  \int_{0}^{1} \!\! dx
   \frac{N_f}{Q^{2}x + q^{2}\bar{x}}~\varphi^{(2)}_{\pi}(x)
   + O \! \left( Q^{-4} \right)&&
\label{Eq:T_0}
\end{eqnarray}
with $N_f=\frac{\sqrt{2}}{3}f_{\pi}$ and $\bar{x}\equiv 1-x$.
The pion DA $\varphi^{(2)}_{\pi}$ is defined as
\begin{eqnarray}
  \langle
         0|\bar{q}(z)\gamma_{\mu}\gamma_{5}\mathcal{C}(z,0) q(0)|\pi(P)
  \rangle
  \Big|_{z^2=0}~=&& \nonumber\\
  ~iP_{\mu}\int dx
  e^{ix(z\cdot p)}\varphi^{(2)}_\pi(x,\mu^2_{\rm F}) \ ,&&
\label{eq:pion-DA}
\end{eqnarray}
where
$
 \mathcal{C}(z,0)
=
 \mathcal{P}\exp \left(ig\int^z_0 A_\mu(\tau) d\tau^\mu \right)
$ is a path-ordered exponential to ensure gauge invariance,
and can be reconstructed from its moments, e.g., within the framework
of QCD sum rules using either local \cite{CZ84} or nonlocal condensates
\cite{MR89,BMS01}.

The radiative corrections to the hard amplitudes $T$ [Eq.\
(\ref{eq:convolution})] in next-to-leading order (NLO), encapsulated in
$T_1$, have been computed in \cite{DaCh81}.
More recently, the $\beta$--part of the next-to-next-to-leading order
(NNLO) amplitude $T_2$, i.e., $\beta_0 \cdot T_{\beta}$, was also
calculated \cite{MMP02}.

\section{Light Cone Sum Rules for the process
$\mathbf{\gamma^*(Q^2)\gamma(q^2\simeq 0) \to \pi^0}$}
\label{sec:LCSR}

Experimentally, the transition form factor was measured by different
Collaborations \cite{CELLO91,CLEO98,BaBar09} when one photon is quasi
real ($q^2 \to 0$).
This kinematics requires the modification of the standard
factorization formula Eq.\ (\ref{eq:convolution}) in order to take
into account the long-distance interaction, i.e., the hadronic content,
of the on-shell photon.
To this end, Khodjamirian, \cite{Kho99}, suggested a light-cone
sum-rule (LCSR) approach, based on a dispersion relation for
$F^{\gamma^{*}\gamma^{*}\pi}$ in the variable $q^2$:
\begin{equation}
  F^{\gamma^{*}\gamma^{*}\pi}\left(Q^2,q^2\right)
=
  \int_{0}^{\infty} ds
  \frac{\rho\left(Q^2,s\right)}{s+q^2}
\label{eq:dis-rel}
\end{equation}
The key element of the LCSR is the spectral density
$\rho(Q^2,s)= \frac{\mathbf{Im}}{\pi}
  \left[F^{\gamma^*\gamma^*\pi}(Q^2,-s)
  \right]
$
for which one may employ the following ansatz \cite{Kho99}:
$\rho= \rho^{\rm ph}(Q^2,s) \theta(s_0-s)
  + \rho^{\rm PT}(Q^2,s) \theta(s-s_0)$,
where  the ``physical'' spectral density $\rho^{\rm ph}$
accumulates the hadronic content of the photon (below the effective
threshold $s_0$) in terms of the form factors of vector mesons, viz.,
\begin{equation}
  \rho^{\rm ph}(Q^2,s)
=
  \sqrt{2}f_V F^{\gamma^*V \pi}(Q^2)
  \cdot
  \delta(s-m^2_{V}) \ ,
\label{eq:resonance}
\end{equation}
where $V$ stands for a $\rho$ or an $\omega$ meson,
while $\rho^{\rm PT}$ contains the partonic part and is
based on Eq.\ (\ref{eq:convolution}) via the relation
$
 \rho^{\rm PT}(Q^2,s)
=
 \frac{\mathbf{Im}}{\pi}
 \left[\left(T \otimes \varphi_{\pi}\right)(Q^2,-s)
  \right]
$.
Using quark-hadron duality in the vector channel, it is possible
to express the pion-photon transition form factor
$F^{\gamma^{*}\gamma\pi}(Q^2,0)$
in terms of $\rho^{\rm PT}$:
\begin{eqnarray}
  F^{\gamma\gamma^*\pi}(Q^2)
=
  \frac{1}{\pi}\int_{s_0}^{\infty}%
  \frac{\textbf{Im}\left(T \otimes \varphi_{\pi}\right)(Q^2,-s)}{s} ds
  \nonumber
\!\!\!&& \\  %
+
  \frac{1}{\pi}\int_{0}^{s_0}%
  \frac{\textbf{Im}\left(T \otimes \varphi_{\pi}\right)(Q^2,-s)}%
  {m_\rho^2}
            e^{(m_\rho^2-s)/M^2}ds . &&
\label{eq:LCSR}
\end{eqnarray}
Here $s_0 \simeq 1.5$~GeV$^2$ and $M^2$ denotes the Borel parameter
in the interval ($0.5-0.9$)~GeV$^2$.

Partial results for $\rho^{(1)}$ at the NLO level have been
presented in \cite{SchmYa99}, while the general solution
$
 \rho_n^{(1)}(Q^2,s)
 =
 \frac{\mathbf{Im}}{\pi}
                        \left[\left(T_{1}\otimes \psi_n\right)(Q^2,-s)
                        \right]
$
was recently obtained in \cite{MS09}:
\begin{eqnarray}
   \bar{\rho}^{(1)}_n\left(x;\mu^2_{\rm F}\right)
=
   C_{\rm F} \left\{
                    -3\left[1-v^{a}(n)\right]+\frac{\pi^2}{3}
                    \right.&& \nonumber \\
         \left.
          -\ln^2\left(\frac{\bar{x}}{x}\right) + 2v(n)
                    \ln\left(\frac{\bar{x}}{x} \frac{Q^2}{\mu^2_{\rm F}}
                       \right)
           \right\} \psi_n(x)
&&  \nonumber \\
    - C_{\rm F}\ 2\!\!
                 \sum^n_{l=0,2,\ldots}\left(G_{nl}
                 +v(n)\cdot b_{n l}\right)\psi_l(x) \ .
\label{eq:spec-den-NLO} &&
\end{eqnarray}
Here $\{\psi_n\}$ are the Gegenbauer harmonics which constitute the
LO eigenfunctions of the Efremov-Radyushkin-Brodsky-Lepage (ERBL)
evolution equation, with $v(n),v^{a}(n)$ being the corresponding
eigenvalues, whereas $G_{nl},~b_{n l}$ are calculable triangular
matrices (see for details \cite{MS09}).

Predictions for $F^{\gamma\gamma^*\pi}(Q^2)$ at the NLO were
given in \cite{BMS02} for various pion DAs, notably for the
asymptotic one, the CZ model \cite{CZ84}, and the BMS model
\cite{BMS01}.
We found that the result of the NLO processing of the CLEO data
\cite{CLEO98}, the BMS bunch of DAs \cite{BMS01}, and the most recent
lattice estimates of the second moment of the pion DA are in good
mutual agreement and inside the $1\sigma$ error ellipse.
The inclusion of the NNLO contribution to the main partial spectral
density $\rho^{(2)}_0$, proportional to $\beta_0$, was realized in
\cite{MS09} taking recourse to the results of \cite{MMP02}.
It turns out that it is negative and about --7\% \footnote{taken
together with the effect of a more realistic Breit-Wigner ansatz
for the meson resonance in Eq.\ (\ref{eq:resonance})} at small
$Q^2\sim 2$~GeV$^2$, decreasing rapidly to --2.5\% at
$Q^2 \geq 6$~GeV$^2$.
The net result is a slight suppression of the prediction for the
scaled form factor (see Fig.\ \ref{fig:NNLO-BaBar}).

\section{NNLO LCSR results vs. BaBar data}
\label{sec:NNLO-BaBar}

\begin{figure}[ht]
\includegraphics[width=0.48\textwidth]{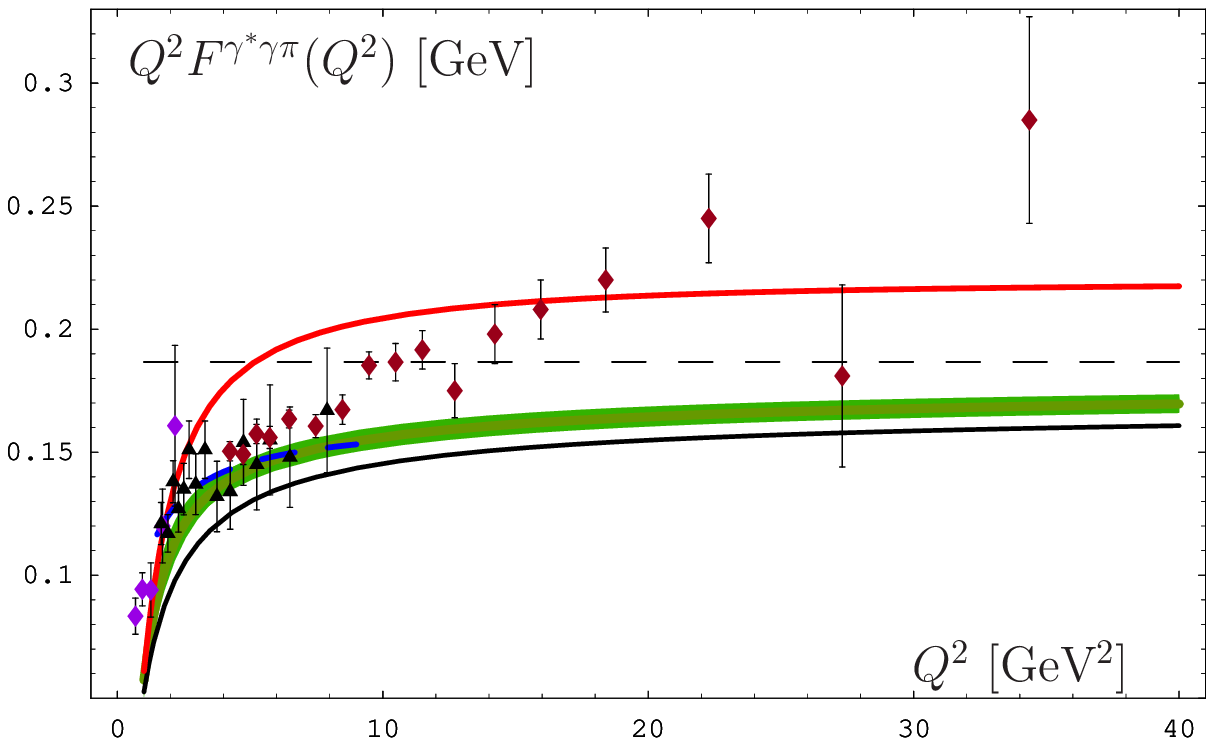}
 \vspace*{-10mm}
   \caption{\footnotesize
 $Q^2F^{\gamma^{*}\gamma\pi}(Q^2)$ calculated with
 three different pion DAs: Asymptotic (lower solid line), BMS
 (shaded green strip), and CZ (upper solid red line).
 The BaBar data \protect\cite{BaBar09} are shown as diamonds
 with error bars.
 The CELLO \protect\cite{CELLO91} and the CLEO data
 \protect\cite{CLEO98} are also shown.
 The displayed theoretical results include the NNLO$_{\beta}$
 radiative corrections and the twist-four contributions.
 The horizontal dashed line marks the asymptotic QCD
 prediction $\sqrt{2}f_\pi$.}
\label{fig:NNLO-BaBar}
\end{figure}

Very recently, the BaBar Collaboration published new results on the
pion-photon transition form factor that cover a wide range of momenta
$4 < Q^2 < 40$~GeV$^2$ with high precision \cite{BaBar09}.
Surprisingly, their data exceed the asymptotic QCD prediction
$\sqrt{2}f_{\pi}$ already at $\sim 10$~GeV$^2$ and continue to
grow with $Q^2$ up to the highest measured momentum.
This behavior would indicate that the $\gamma^*\gamma\to\pi$ process
cannot be correctly described within the convolution scheme of QCD
based on the collinear factorization.
We argued in \cite{MS09} that the inclusion of the NNLO radiative
corrections cannot reconcile the BaBar data with perturbative
QCD for any pion DA that vanishes at the endpoints $x=0,1$---see
Fig.\ \ref{fig:NNLO-BaBar}.
Indeed, from Table \ref{table:1} it becomes clear that also a wide
pion DA, like the CZ model, cannot reproduce \emph{all} BaBar data
both in isolation or jointly with the CLEO data \cite{CLEO98}.
We conclude: \\
(i) The combined effect of the negative ${\rm NNLO}_\beta$ radiative
corrections and the Breit-Wigner ansatz for the meson resonances in
Eq.\ (\ref{eq:resonance}) results in a moderate suppression of
$Q^2F^{\gamma^{*}\gamma\pi}(Q^2)$ in the range of momentum transfers
10-40~GeV$^2$ \cite{MS09}. \\
(ii) The hadronic content of the real photon is a twist-four
contribution that is rapidly decreasing with increasing $Q^2$ and can,
therefore, not be the origin of the enhancement of the scaled form
factor measured by BaBar. \\
(iii) Within the QCD convolution scheme, \emph{all} pion DA models
(cf.\ the $\chi^2$ in Table \ref{table:1}), which have a convergent
projection onto the Gegenbauer harmonics, and hence vanish at the
endpoints 0, 1, are in conflict with the BaBar data for
$Q^2F^{\gamma^{*}\gamma\pi}(Q^2)$ between 10 and 40~GeV$^2$
(see Fig.\ \ref{fig:NNLO-BaBar}), because in this range these data
violate the collinear factorization formula per se.
\vspace*{-3mm}
\hspace*{-5mm}
\begin{table}[ht]
 \caption{ \footnotesize $\chi^2_{ndf}$ for Asymptotic (Asy), BMS, and CZ DAs}
 \label{table:1}
\begin{tabular}{p{0.07\textwidth}|p{0.07\textwidth}|p{0.07\textwidth}|p{0.13\textwidth}} \hline
   Pion DA             & BaBar and          &  BaBar        & BaBar only 10 data with \\
                       & CLEO               &  all~data     &   $Q^2~>~10$~GeV$^2$ \\    \hline
 Asy           & $11.5$   ~~~           & $19.2$      ~~             &$19.8$ ~~ \\
 BMS   & $4.4$~~~&$ 7.8$~~    & $11.9$ ~~ \\
 CZ       & $20.9$   ~& $36.0$     ~~& $6.0$  ~~ \\                                      \hline
\end{tabular}
\end{table}

\section{Can the BaBar data be explained?}
\label{sec:BaBar-scenarios}

\begin{figure}[hbt]
  \includegraphics[width=0.48\textwidth]{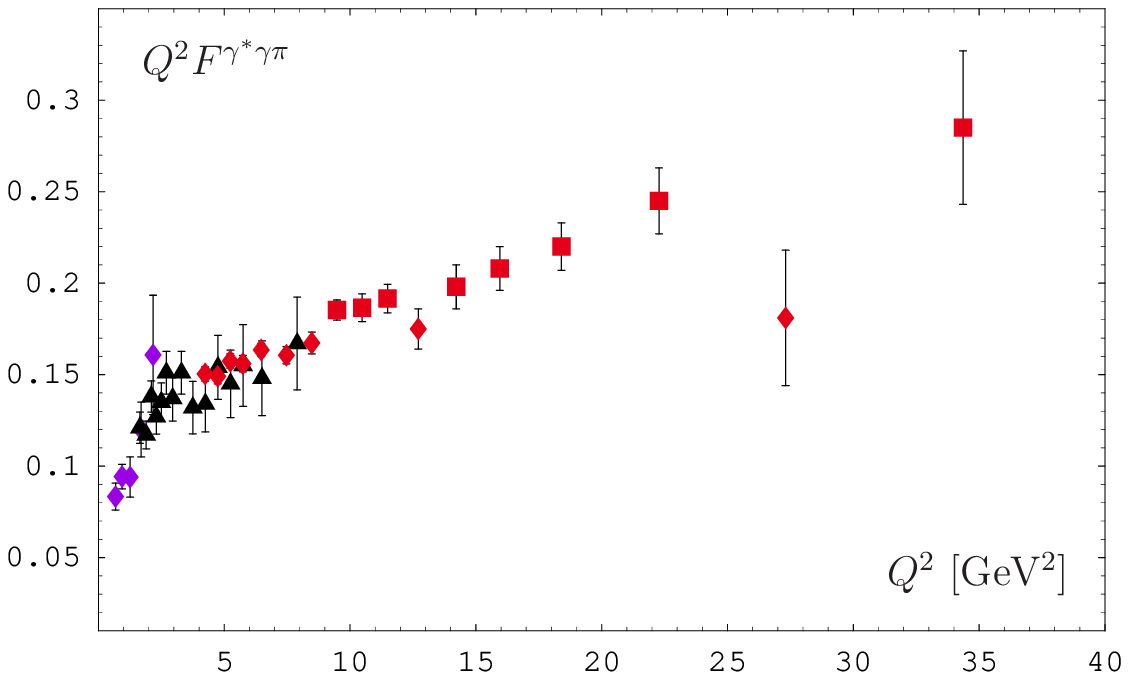}
   \vspace*{-10mm}
   \caption{\footnotesize Split BaBar data:
   \RedTn{\footnotesize \ding{117}} - BaBar1,
   \RedTn{\footnotesize\ding{110}} - BaBar2.}
   \label{fig:split-BaBar}
 \end{figure}

There are no formal explanations of the BaBar data within the general
framework of QCD at present---there are no reasons to adduce.
However, some theoretical scenarios have already been proffered
\cite{Dor09,Rad09,Pol09}.
We will discuss one class of such proposals based on the idea that
the pion DA may be ``practically flat'', hence violating the
collinear factorization and entailing a (logarithmic) growth of
$Q^2F^{\gamma^*\gamma\pi}(Q^2)$ with $Q^2$.
One has \cite{Rad09}
($\sigma =0.53$~GeV$^2$; $\varphi_{\pi}(x)=f_{\pi}$)
\begin{equation}
  Q^{2}F^{\gamma^{*}\gamma\pi}
=
  \frac{\sqrt{2}}{3}
  \int_{0}^{1} \frac{\varphi_{\pi}(x)}{x}
 \Ds \left[1 - {\rm e}^{ -\frac{x Q^2}{\bar{x}2 \sigma}}\right]dx \ .
\label{eq:flat-Rad}
\end{equation}
Another option \cite{Pol09} gives instead ($m\approx 0.65$~GeV)
\begin{equation}
 Q^{2}F^{\gamma^{*}\gamma\pi}
=
  \frac{\sqrt{2}}{3}
  \int_{0}^{1}
  \frac{\varphi_{\pi}(x,Q)}{x + \frac{m^2}{Q^2}}dx\,
\label{eq:flat-Pol}
\end{equation}
with $\varphi_{\pi}(x,\mu_0)=f_{\pi}(N+(1-N)6x\bar{x}),~(N\approx 1.3)$.
Eq. (\ref{eq:flat-Rad}) can be compared with the available
experimental data \cite{CELLO91,CLEO98,BaBar09} by parameterizing
them via the phenomenological fit ($\Lambda\approx 0.9$~GeV, $b\approx-1.4$)
$$
  Q^{2}F^{\gamma^{*}\gamma\pi}
= \!
  \frac{Q^2}{2 \sqrt{2} f_{\pi} \pi^2}
  \!\left[\frac{\Lambda^2}{\Lambda^2+Q^2}+
  b \left(\!\!\frac{\Lambda^2}{\Lambda^2+Q^2} \!\!\right)^2 \right]
\label{eq:dipole}
$$
as shown in Table \ref{table:dipole-flat}.
The main message is that one cannot fit the CELLO/CLEO data
and the BaBar data with the same accuracy simultaneously
with both parameterizations.
 \begin{table}[ht]
  \caption{\footnotesize $\chi^2_{ndf}$ for the phenomenological fit
  and the fit with a flat DA like Eq.\
  \protect(\ref{eq:flat-Rad})---numbers in parentheses.
  The diagonal elements give the best-fit values of
  $\chi^2_{ndf}$ which fix the corresponding line parameters.}
\begin{tabular}{p{0.1\textwidth}|p{0.1\textwidth}|p{0.1\textwidth}}
               & \hspace*{-2mm} CELLO\& \hspace*{2mm} CLEO
               & \hspace*{-2mm} BaBar  \\  \hline
 \hspace*{-2mm}CELLO\& \hspace*{-2mm}CLEO
               &  $0.48~(1.22)$  & $7.8~(15.8)$      \\ \hline
 \hspace*{-2mm}BaBar$\vphantom{^\big|_|}$
               &  $10.8~(3.5)$        & $1.8~(1.8)$  \\ \hline
\label{table:dipole-flat}
\end{tabular}
\end{table}
Now let us push this point further and divide the BaBar data into
two `experiments' BaBar1 and BaBar2, as indicated graphically in
Fig.\ \ref{fig:split-BaBar}.
One sees from Table \ref{table:split-BaBar} in terms of a $\chi^2_{ndf}$
criterion that the flat-DA scenario cannot describe both BaBar
`experiments' simultaneously with the same accuracy.
One could interpret this outcome as an indication for an intrinsic
inconsistency in the analysis of the BaBar data that deserves further
attention.

\begin{table}[htb]
\caption{\footnotesize $\chi^2_{ndf}$ for the flat-DA fit. Best-fit
values on the diagonal.}
\label{table:split-BaBar}
\begin{tabular}{p{0.1\textwidth}|p{0.1\textwidth}|p{0.1\textwidth}}
               & \RedTn{\footnotesize \ding{117}} BaBar1
               & \RedTn{\footnotesize\ding{110}}  BaBar2 \\  \hline
 \hspace*{-2mm}BaBar1
               & $3.3$    & $0.33$      \\  \hline
 \hspace*{-2mm}BaBar2$\vphantom{^\big|_|}$
               & \vspace*{-2mm}$3.5$
               & \vspace*{-2mm}$0.26$    \\ \hline
\end{tabular}
\end{table}

\vspace*{-3mm}
\section{Conclusions}
\label{sec:concl}
In conclusion, we have presented a calculation within the LCSR approach
of the pion-photon transition form factor which includes the main NNLO
radiative corrections and twist-four contributions.
Our predictions are based on collinear factorization and agree with the
data of the CELLO and the CLEO Collaborations but greatly disagree
with the new high-$Q^2$ BaBar data.
Our analysis shows that all pion DAs, which vanish at the endpoints
$x=0,1$, cannot reproduce the observed growth of the scaled form factor
above 10~GeV$^2$.
On the other hand, flat pion DAs may describe this growth at high $Q^2$,
but at the expense that they fail to comply with the whole set of the
BaBar data and also with those of the CELLO and the CLEO Collaborations.
Future experiments may clarify this situation.

\section*{Acknowledgments}
This work was partially supported by the Heisenberg--Landau
Program (Grant 2009) and the Russian Foundation for Fundamental
Research (Grants 07-02-91557 and 09-02-01149).

\end{document}